\documentclass[preprint,review,12pt]{elsarticle}
\newcommand{\alumina}{Al$_2$O$_3$}

\makeatletter
\def\ps@pprintTitle{%
  \let\@oddhead\@empty
  \let\@evenhead\@empty
  \def\@oddfoot{\reset@font\hfil\thepage\hfil}
  \let\@evenfoot\@oddfoot
}
\makeatother

\usepackage{graphicx}
\usepackage{lineno}
\usepackage{upgreek}

\begin{document}

\linenumbers

\begin{frontmatter}

\title{Efficient calculation of local dose distribution for response modelling in proton and ion beams}

\author[label1]{Greilich, S.\corref{cor1}}
\ead{s.greilich@dkfz.de}
\author[label2]{Hahn, U.}
\author[label2]{Kiderlen, M.}
\author[label3]{Andersen, C.E.}
\author[label4]{Bassler, N.}

\address[label1]{Department of Medical Physics in Radiation Oncology, German Cancer Research Center (DKFZ), Im Neuenheimer Feld 280, D-69120 Heidelberg, Germany}
\address[label2]{Department of Mathematical Sciences, Aarhus University, Ny Munkegade, bygning 1530, DK-8000 Aarhus C, Denmark}
\address[label3]{Center for Nuclear Technologies, Technical University of Denmark, Ris{\o} Campus, P.O. 49, DK-4000 Roskilde, Denmark}
\address[label4]{Department of Experimental Clinical Oncology, Aarhus University Hospital, N{\o}rrebrogade 44, DK-8000 Aarhus C, Denmark}

\cortext[cor1]{Corresponding author, Tel: +49-(0)6221-42-2632, Fax: +49-(0)6221-42-2665}


\begin{abstract}
We present an algorithm for fast and accurate computation of the local dose distribution in MeV beams of protons, carbon ions or other heavy-charged particles. It uses compound Poisson-process modelling of track interaction and succesive convolutions for fast computation. It can handle mixed particle fields over a wide range of fluences. Since the local dose distribution is the essential part of several approaches to model detector efficiency or cellular response it has potential use in ion-beam dosimetry and radiotherapy.
\end{abstract}

\end{frontmatter}

\section{Introduction}

\begin{figure}
	\centering
		\includegraphics[width = \textwidth]{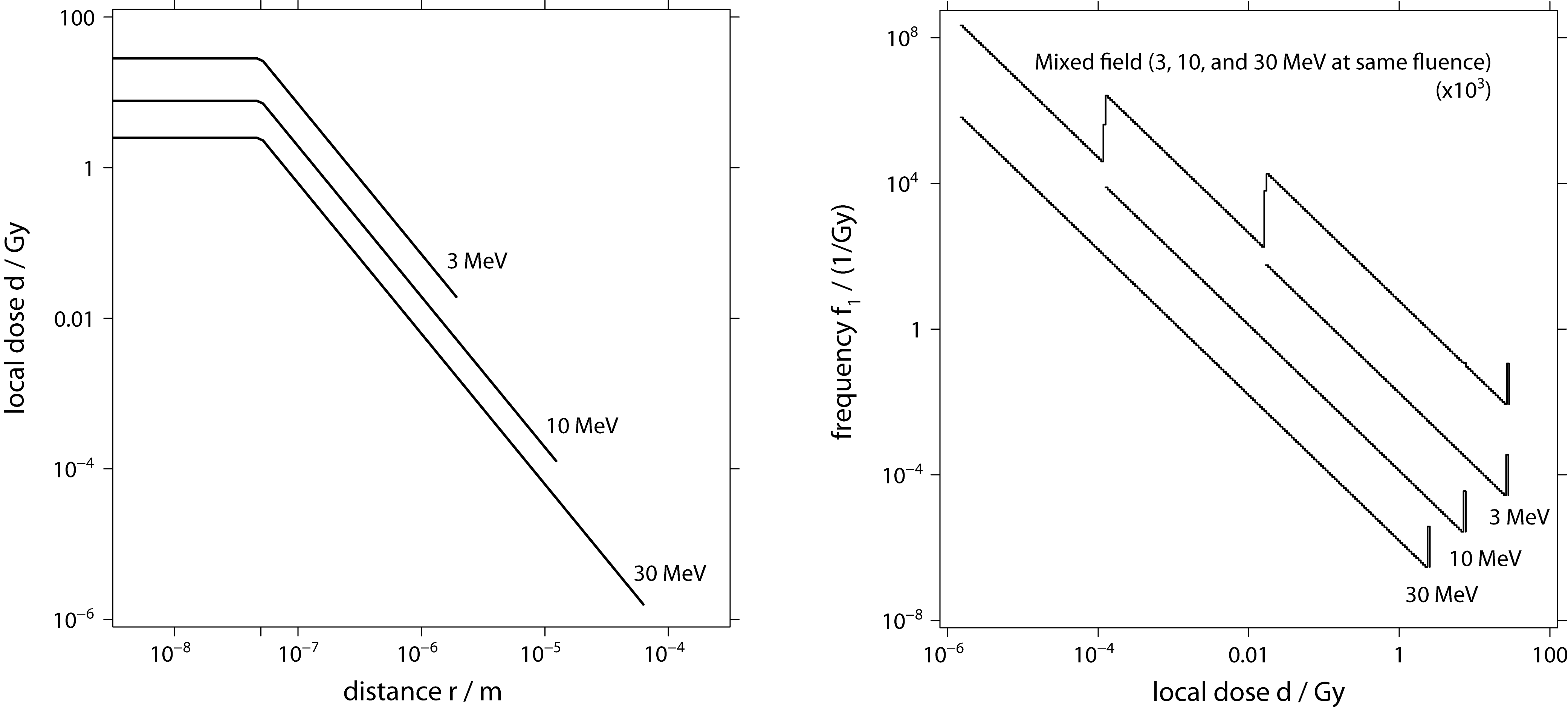}
	\caption{Left: Radial dose distributions $d(r)$ in water for three proton energies. Considerable differences exist between parameterizations from different authors. For simplicity, we use the one by Scholz as in \cite{Edmund_et_al_2007}. Right: Single impact distributions $f_1(d)$ for the same particles plus a mixed field case (for visibility scaled by $10^3$) with logarithmic binning (10 bins per factor of 10). Although relatively unlikely, high local doses in the core region can contribute significantly to the total dose (tens of percent) and must not be neglected.}
		\label{fig:Figure1}
\end{figure}

Amorphous track models (ATMs) disregard the stochastic energy deposition pattern by secondary electrons around the track of densily ionizing, heavy charged particles (i.e. protons or ions, HCPs). Rather, they consider an dose average as a two-dimensional, radially symetric function of perpendicular distance $r$ from the trajectory, referred to as radial dose distribution $d(r)$ (\mbox{Fig. \ref{fig:Figure1}}). 

In contrast, sparsely ionizing radiation such as photons deposits energy by secondary electrons as well, yet ionization events are supposed to be homogeneously distributed across the irradiated media. The 'local effect' assumption in ATMs assumes that on small scales ($\ll \mathrm{\upmu m}$) the energy deposition from HCPs cannot be distinguished from a photon field.

The response to irradiation with HCPs of type $T$ and energy $E$ can therefore be predicted from the homogenous bulk photon dose response $S_X(D)$ of the system and the spatial deposition of local dose $d(x,y)$ as calculated from the fluences $\Phi(E, T)$ of the particle field. Despite many simplifications, ATMs are reasonably successful in predicting the response for a variety of physical detectors and biological systems \cite{Katz_et_al_1972, Waligorski_and_Katz_1980, Geiss1998, Bassler_et_al_2008}. In a former study \cite{Andersen_et_al_2009}, we employed ATMs for our all-optical dose verification system using fiber-coupled \alumina :C as prerequisit for use in particle beams \cite{Edmund_et_al_2007, Greilich_et_al_2008}.

In \cite{Greilich_et_al_2008} we used a simple generic grid summation (GSM) based on a Monte-Carlo technique. There, particles are sampled according to their relative fluence and local doses $d(x,y)$ --- and therefore local response $s(x,y)=S_X(d(x,y))$ --- are computed on a Cartesian grid ('checkerboard') by attributing the corresponding $d_{E,T}(r)$ to the sampled particle (\mbox{Fig. \ref{fig:Figure2}}). The detector is thought to be homogenous, perpendicular to the beam in $(x,y)$, and of negligible thickness $\Delta z$. The relative efficiency $\eta$ can then be estimated by averaging the local response $s$ over all grid elements:
\begin{equation}\label{eq:eta}
	\eta(\phi(E, T))=\frac{S_{HCP}}{S_X(D)}=\frac{\left\langle s\right\rangle}{S_X(\left\langle d\right\rangle)}
\end{equation}

Although conceptually straightforward, GSM can be very time-consuming, esp. in the case of higher fluences and particle energies (e.g. $E_{\mbox{proton}} >$ 20 MeV) with many contributions to a single voxel. Furthermore, the procedure has to be repeated many times in order to converge --- or a large detector grid has to be simulated.

\section{Computation of the local dose distribution using compound Poisson processes}

\begin{figure}
	\centering
		\includegraphics[width = \textwidth]{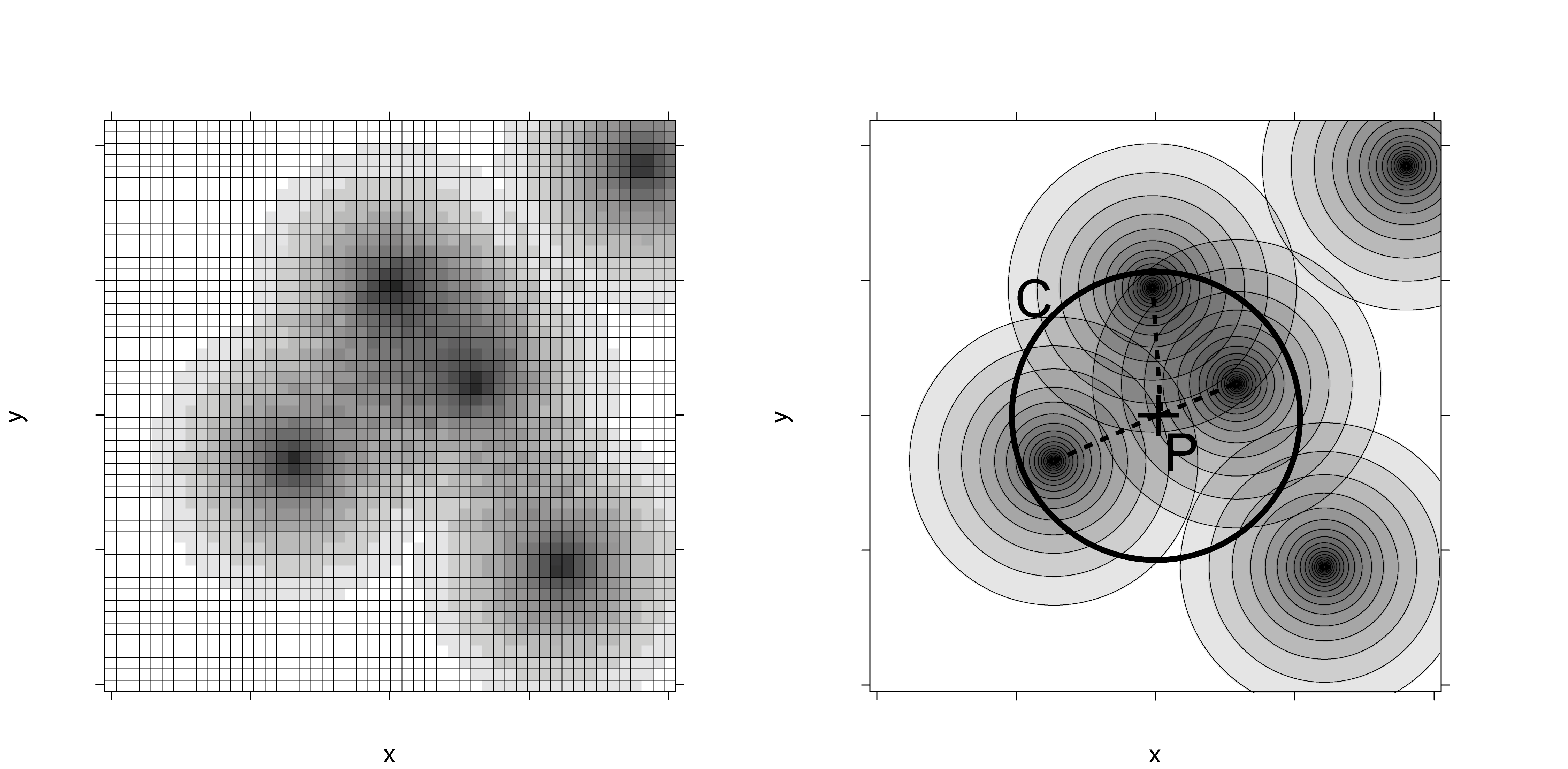}
	\caption{Illustration of the GSM (left) and the compound Poisson modelling.}
		\label{fig:Figure2}
\end{figure}

To overcome the limitations of GSM calculating the local dose distibution as a spatial deposition pattern $d(x,y)$, we consider a representative point $P$ (Fig. \ref{fig:Figure2}). The cumulative distribution function $F(d)$ of local dose $d$ in $P$ depends on the macroscopic fluence $\phi$ (and dose $D$, resp.) and the microscopic pattern around a track as expressed by $d(r)$. Then, one can state:
\begin{itemize}
	\item {$r_{max}$ is the maximum seondary electron range in the field, so $P$ is only influenced by tracks within a circle $C$ of radius $r_{max}$ around $P$ (\mbox{Fig. \ref{fig:Figure2}}).}
	\item {All tracks in $C$ are contributing to $d$ and their number $n$ is Poisson distributed with mean $\mu=\phi\cdot\pi r_{max}^2$.}
	\item {Let $F_n$ be the cumulative distribution function of the local dose in the case of exactly $n$ tracks. For a single track traversing $C$, we readily have the cumulative single impact distribution
		\begin{equation}\label{eq:singletrack}
			F_1(d)=1-\frac{R(d)^2}{r_{max}^2},
		\end{equation}
		with $R(d)=D^{-1}(r)$ (\mbox{Fig. \ref{fig:Figure1}}).}
	\item {In the case of $n$ tracks in $C$, $d$ is the sum of $n$ independent and identically distributed single track doses, so $F_n$ can be expressed as the $n$-fold convolution of $F_1$:
		\begin{equation}\label{eq:ntracks}
			F_n=\underbrace{F_1*\ldots*F_1}_{\mbox{$n$ times}}
		\end{equation}
	}
	\item {As $n$ is Poisson distributed, $F$ is the distribution function of a compound Poisson process:
		\begin{equation}
			F(d)=e^{-\mu}\sum^{\infty}_{i=1}{\frac{\mu^i}{i!}F_i(d)}
		\end{equation}
	}
	\item {The derivative $f(d)$ of $F(d)$ can then be eventually used to compute the macroscopic HCP response as the expected local response $\left\langle s\right\rangle$:
		\begin{equation}
		  \langle s \rangle= \int_0^{\max(d)} S_X(d) f(d)\, \mathrm{d}d,
		\end{equation}
		and used in Eq. (\ref{eq:eta}) to get the $\eta$. A similar procedure for $\left\langle d\right\rangle$ provides a quality check as it has to meet $D$.
	}
\end{itemize}
This description enables $F$ to be determined from the explicitly given distribution function $F_1$ in the case of monoenergetic particle fields. It can easily be extended to mixed particle fields by using the adjusted $F_1$ from Eq.(\ref{eq:multitrack}) in Eq.(\ref{eq:ntracks}) with $p_{E,T}$ being the relative fluence and $R_{E,T}$ the inverse radial dose distribution for the composing particles
\begin{equation}\label{eq:multitrack}
  F_1(d)=1-\sum_{E,T}p_{E,T}\cdot\frac{R_{E,T}(d)^2}{\hat{r}_{max}^2},
\end{equation}
where $\hat{r}_{max}=max(r_{max}(E,T))$.

It should be stressed that the presented approach is in no way limited to handle extended targets despite the point nature of $P$ as the averaging across the target is already contained in $D(r)$ (for the difference between point and extended target distributions, see \cite{Katz_et_al_1972, Edmund_et_al_2007}). While the computation of detector response from the local dose distribution $F(d)$ is trivial, the numerical calculation of $F(d)$ itself can, however, be cumbersome.

\section{Accelerated computation using successive convolution}

An approximation method for the rapid computation of compound Poisson processes was introduced by Kellerer \cite{Kellerer_1985}. It makes use of the fact that the distribution $f(d;\mu)$ can be obtained by a convolution operation:
	\begin{equation}
		f(d;\mu)=\int_{0}^{d}{f(d-t;\mu/2)\cdot f(t;\mu/2) dt}
	\end{equation}
One can chose a $\mu_{start}<<1$ with $\mu=2^m\cdot\mu_{start}$ so that multiple events can be neglected and therefore $f(d;\mu_{start})$ consists, in good approximation, of two components only, namely the probability of no track in $C$ ($d=0$) 
	\begin{equation}
		e^{-\mu_{start}}\approx (1-\mu_{start})=\hat{f}_0
	\end{equation}
and of the density related to a single track
	\begin{equation}
		\hat{f_1}\approx\mu_{start}\cdot f_1(d).
	\end{equation}
Performing $m$ successive convolutions on these two components, i.e. replacing
	\begin{equation}
		\hat{f}_0 \mbox{ by } \hat{f}_0^2
	\end{equation}
and
	\begin{equation}
		\hat{f}_1 \mbox{ by } 2\cdot\hat{f}_0\cdot\hat{f}_1 + \hat{f}_1*\hat{f}_1
	\end{equation}
will eventually yield $f(d)$.

\section{Results and conclusion}
\begin{figure}
	\centering
		\includegraphics[width = \textwidth]{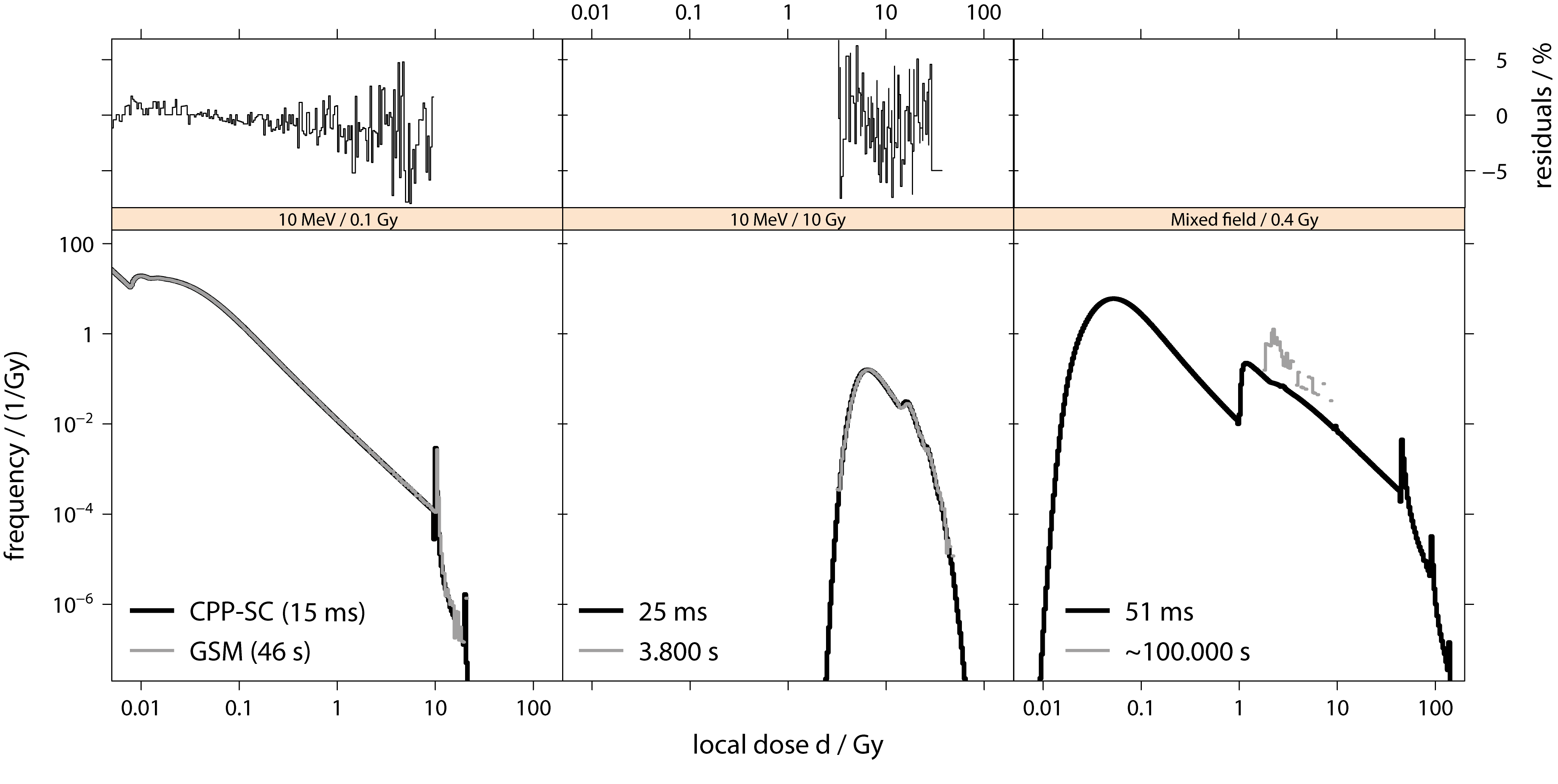}
	\caption{Comparison of the resulting local dose distribution $f(d)$ from GSM and the algorithm based on compound Poisson processes using successive convolutions (CPP-SC). Left: At low fluence the structure of $f_1(d)$ still visible, but a low dose dip and multiple core events occur due to track overlap. Middle: High fluence case. With many contribution tracks, $f(d)$ approaches a narrow Gaussian profile corresponding to a homogenous dose over the detector ('sea of electrons'). Right: Mixed field case (as in Fig. \ref{fig:Figure1}). For GSM, normalization fails due to edge effects.}
		\label{fig:Figure3}
\end{figure}

Fig. \ref{fig:Figure3} shows that the new algorithm delivers $f(d)$ --- as the essential part of modelling detector efficiency and cellular response in HCP beams --- in very good agreement with GSM but significantly faster. In addition, it covers a much wider range of local dose and can handle high energy (i.e. very wide) tracks like the 30 MeV component in the mixed field that clearly overstrains the capabilities of GSM. The novel algorithm presented thus allows a much larger parameter space in ATM modelling. Using it, spatial binning becomes obsolete. In addition, even complex particle fields can easily be treated as they mainly affect the computation of $F_1$, the number of convolutions. We believe that the algorithm can contribute to the applicability of ATMs and the improvement of accuracy in their predictions for HCP dosimetry and radiotherapy. It has been implemented as part of the open-source project libamtrack (http://libamtrack.dkfz.org).

\section*{Acknowledgments}
The authors gratefully acknowledge the support by CIRRO (The Lundbeck Foundation Center for Interventional Research in Radiation Oncology) and The Danish Council for Strategic Research. NB acknowledges support from the Danish Cancer Society and the German Research Foundation (DFG).

\section*{References}
\bibliography{spiff}

\begin{thebibliography}{1}

\bibitem{Katz_et_al_1972}
R.~Katz, S.~C. Sharma, and M.~Homayoonfar.
\newblock The structure of particle tracks.
\newblock In F.~H. Attix, editor, {\em Topics in Radiation Dosimetry},
  chapter~6, pages 317--383. Academic Press, New York, 1972.

\bibitem{Waligorski_and_Katz_1980}
M.P.R. Walig\'orski and R.~Katz.
\newblock {Supralinearity of peak 5 and peak 6 in TLD-700}.
\newblock {\em NIM}, 172:463--470, 1980.

\bibitem{Geiss1998}
O.~B. Gei{\ss}, M.~Kr\"{a}mer, and G.~Kraft.
\newblock Efficiency of thermoluminescence detectors to heavy charged
  particles.
\newblock {\em NIM B}, 142:592--598, 1998.

\bibitem{Bassler_et_al_2008}
N.~Bassler, J.W. Hansen, H.~Palmans, M.H. Holzscheiter, S.~Kovacevic, and the
  AD-4/ACE~Collaboration.
\newblock {The antiproton depth–dose curve measured with alanine detectors}.
\newblock {\em NIM B}, 266:929--936, 2008.

\bibitem{Andersen_et_al_2009}
C.E. Andersen, S.K. Nielsen, S.~Greilich, J.~Helt-Hansen, J.C. Lindegaard, and
  K.~Tanderup.
\newblock {Characterization of a fiber-coupled \alumina :C luminescence
  dosimetry system for online in vivo dose verification during 192Ir
  brachytherapy}.
\newblock {\em Med. Phys.}, 36:708--718, 2009.

\bibitem{Edmund_et_al_2007}
J.~Edmund, C.~Andersen, and S.~Greilich.
\newblock {A track structure model of optically stimulated luminescence from
  \alumina :C irradiated with 10-–60 MeV protons}.
\newblock {\em NIM B}, 21:261--275, 2007.

\bibitem{Greilich_et_al_2008}
S.~Greilich, J.M. Edmund, M.~Jain, and C.E. Andersen.
\newblock {A coupled RL and transport model for mixed-field proton irradiation
  of \alumina :C}.
\newblock {\em Rad. Meas.}, 43:1049–1053, 2008.

\bibitem{Kellerer_1985}
A.M. Kellerer.
\newblock Fundamentals of microdosimetry.
\newblock In K.R. Kase, B.E. Bj\"arngard, and F.H. Attix, editors, {\em The
  Dosimetry of Ionizing Radiation}, chapter~2. Academic Press, London, 1985.

\end{thebibliography}

\newpage

\newpage

\newpage

\end{document}